\documentclass[conference]{IEEEtran}
\IEEEoverridecommandlockouts
\usepackage{comment}
\usepackage{multirow, array}
\usepackage{cite}
\usepackage{eurosym}
\usepackage[mathscr]{euscript}
\usepackage{amsmath,amssymb,amsfonts}
\usepackage{algorithmic}
\usepackage{graphicx}
\usepackage{textcomp}
\usepackage{xcolor}
\def\BibTeX{{\rm B\kern-.05em{\sc i\kern-.025em b}\kern-.08em
    T\kern-.1667em\lower.7ex\hbox{E}\kern-.125emX}}

\newenvironment{ldescription}[1]
  {\begin{list}{}%
   {\renewcommand\makelabel[1]{##1\hfill}%
   \settowidth\labelwidth{\makelabel{#1}}%
   \setlength\leftmargin{\labelwidth}
   \addtolength\leftmargin{\labelsep}}}
  {\end{list}}

\begin{document}

\IEEEoverridecommandlockouts

\title{Day-ahead Operation of an Aggregator of Electric Vehicles via Optimization under Uncertainty
\thanks{This project has received funding from the European Research Council (ERC) under the European Union's Horizon 2020 research and innovation programme (grant agreement No 755705). This project has also been supported by Fundaci\'on Iberdrola Espa\~na 2018.}}

\author{\IEEEauthorblockN{\'Alvaro Porras Cabrera}
\IEEEauthorblockA{\textit{OASYS group} \\
\textit{University of Malaga}\\
Malaga, Spain\\
alvaroporras19@gmail.com}
\and
\IEEEauthorblockN{Ricardo Fern\'andez-Blanco}
\IEEEauthorblockA{\textit{OASYS group} \\
\textit{University of Malaga}\\
Malaga, Spain\\
Ricardo.FCarramolino@uma.es}
\and
\IEEEauthorblockN{Juan Miguel Morales}
\IEEEauthorblockA{\textit{Dept. Applied Mathematics} \\
\textit{University of Malaga}\\
Malaga, Spain \\
Juan.Morales@uma.es}
\and
\IEEEauthorblockN{Salvador Pineda}
\IEEEauthorblockA{\textit{Dept. Electrical Engineering} \\
\textit{University of Malaga}\\
Malaga, Spain \\
spinedamorente@gmail.com}
}

\maketitle

\begin{abstract}
%
We assume an aggregator of electric vehicles (EVs) aiming to buy energy in the day-ahead electricity market while accounting for the technical aspects of each individual EV.
{\color{black} We pose the aggregator's problem as a bilevel model, where the upper level minimizes the total operation costs of the fleet of EVs, while each lower level minimizes the energy available to each vehicle for transportation given a certain charging plan. Thanks to the totally unimodular character of the constraint matrix in the lower-level problems, the model can be mathematically recast as a computationally efficient mixed-integer program that delivers charging schedules that are robust against the uncertain availability of the EVs}. Finally, we use synthetic data from the National Household Travel Survey 2017 to analyze the behavior of the EV aggregator from both economic and technical viewpoints and compare it with the results from a deterministic approach.
\end{abstract}
{\color{black}
	}
	
\vspace{-0.2cm}
\section{Nomenclature}
\vspace{-0.2cm}
\subsection{Sets and Indices}
\begin{ldescription}{$xxx$}
\item [$\mathcal{V}$] Set of electric vehicles, indexed by $v$.
\item [$\mathcal{T}$] Set of time periods, indexed by $t$.
\end{ldescription}
\vspace{-0.2cm}
\subsection{Variables}
\begin{ldescription}{$xxx$}
\item [$c_{v,t}$] Charging power of electric vehicle $v$ in period $t$, kW.
\item [$e_{v,t}$] Energy state of charge of vehicle $v$ in period $t$, kWh.
\item [$s^+_{v,t}$] Slack variable for the energy balance of the battery of electric vehicle $v$ in period $t$, kWh.
\item [$s^-_{v,t}$] Slack variable for the energy balance of the battery of electric vehicle $v$ in period $t$, kWh.
\item [$z_{v,t}$] Auxiliary variable used to linearize $c_{v,t} \alpha_{v,t}$, kW.
\item [$\alpha_{v,t}$] Availability of electric vehicle $v$ to charge, being 1 if available and 0 otherwise.
\item [$\xi_{v}^{wc}$] Energy required for transportation of electric vehicle $v$ throughout the optimization horizon for the worst-case scenario.
\item [$\zeta_{v}$] Dual variable associated with the constraint ensuring a minimum number of periods in which the electric vehicle $v$ is available.
\item [$\underline{\beta}_{v,t}$] Dual variable associated with the lower bound on the availability of electric vehicle $v$ and time period $t$.
\item [$\overline{\beta}_{v,t}$] Dual variable associated with the upper bound on the availability of electric vehicle $v$ and time period $t$.
\end{ldescription}
\vspace{-0.2cm}
\subsection{Parameters}
\begin{ldescription}{$xxx$}
\item [$\overline{C}_{v}$] Maximum charging power of electric vehicle $v$, kW.
\item [$\overline{E}_{v}$] Maximum state of charge of electric vehicle $v$, kWh.
\item [$\underline{E}_{v}$] Minimum state of charge of electric vehicle $v$, kWh.
\item [$N_T$] Number of time periods of the optimization horizon.
\item [$P$] Penalty cost, $\textup{\euro}$/kWh.
\item [$\eta_{v}$] Charging efficiency of electric vehicle $v$.
\item [$\Delta t$] Time interval, h.
\item [$\lambda_t$] Day-ahead electricity price in time period $t$, $\textup{\euro}$/kWh.
\item [$\widehat{\alpha}_{v,t}$]  Expected availability of electric vehicle $v$ in period $t$.
\item [$\mathcal{\widehat{\xi}}_{v,t}$] Expected energy consumption of electric vehicle $v$ in period $t$ due to motion, kWh.
\item [$\overline{\alpha}_{v,t}$]  Maximum value of the availability of electric vehicle $v$ in period $t$.
\item [$\underline{\alpha}_{v,t}$] Minimum value of the availability of electric vehicle $v$ in period $t$.
\item [$K_{v}$] Minimum number of periods in which the electric vehicle $v$ must be available throughout the time horizon.
\end{ldescription}

\section{Introduction}
\label{sec:introduction}
According to the Directive 2018/844 of the European Parliament and of the Council of 30 May 2018 \cite{directive_844}, the electrification of transport will be ``\emph{promoted in the near future while enabling further development at a reduced cost in the medium to long term}.'' Thus, the electromobility is expected to impact the operation and planning of future electricity systems and electric vehicles (EVs) will play an important role \cite{bunsen2018global}. Within this context, new actors will appear in the power system, e.g. aggregators, in order to manage the operation of a fleet of EVs in residential and non-residential areas. The successful roll-out of EVs will undoubtedly bring environmental and societal benefits, however such aggregators will face challenges related to their operation and planning.

There is a vast literature on the impact of EVs on the distribution system \cite{de2017impact,fernandez2011assessment,heydt1983impact} and on the charging strategies of a single EV or a fleet of EVs \cite{halvgaard2012electric,iversen2014optimal,sarker2017optimal,ortega2014optimal,baringo2017stochastic,vaya2015integration}. This paper is focused on the latter one, specifically at residential level. It is obvious that a growing penetration of EVs in a residential district will lead to an undesirable over-consumption at certain time periods and the charging of the EVs will be affected due to feeder limitations. In addition, the drivers' habits such as arrival and departure times are stochastic in nature and unpredictable. The treatment of this type of uncertainty is still an unresolved issue.

Halvgaard \emph{et al.} \cite{halvgaard2012electric} implemented an economic model predictive control to simulate the charging strategy of an EV by considering perfect forecasts of the driving patterns. Iversen \emph{et al.}  \cite{iversen2014optimal} handled the uncertainty on the driving patterns for a single EV with an inhomogeneous Markov model. However, upgrading such stochastic dynamic programming model to a fleet of EVs may be complex. Sarker \emph{et al.} \cite{sarker2017optimal} put forward a model for a charging station coupled with a Battery Energy Storage System (BESS) wherein there exist bidirectional communication between the BESS and the grid. In this paper, the technical and physical characteristics of each EV were neglected, and thus the aggregated EV operation was represented by a charging demand of the fleet, which was modelled via stochastic optimization. Ortega-Vazquez \cite{ortega2014optimal} proposed an optimization problem to solve the scheduling of an aggregator at household level while accounting for both the technical characteristics of the EV and their battery degradation costs with an emphasis on vehicle-to-grid services. However, it neglected the stochastic nature of the driving patterns of an EV.  Baringo and S\'anchez Amaro \cite{baringo2017stochastic} analyzed the bidding strategy problem for an EV aggregator participating in the day-ahead electricity market while accounting for the uncertainty on the driving requirements. The EV aggregator was modelled as a virtual battery, thus ignoring the individual operation of each EV of the fleet; and the uncertainty was represented with confidence bounds on the aggregated demand. Similarly, Gonz\'alez Vay\'a \emph{et al.} \cite{vaya2015integration} analyzed the same problem but the uncertainty on driving patterns was represented by means of scenarios.

This paper aims to model the day-ahead operation of a residential aggregator of a fleet of EVs. The technical and physical characteristics of each EV are taken into consideration individually based on the users' information, i.e., the aggregator seeks to optimize the amount of electricity to be purchased in each time period. In addition, the uncertainty on driving patterns is modelled via robust optimization due to the limited information the aggregator will be able to collect. The main contributions of this paper are threefold:

\begin{itemize}
    \item We propose a novel mathematical model to operate a fleet of EVs in a residential area while accounting for the uncertainty on driving patterns. Moreover, the operation of each EV is incorporated into the problem formulation.
    \item We model the uncertainty on driving patterns by using robust optimization so that the total energy required for transportation for each EV must be satisfied throughout the optimization horizon when considering the worst-case scenario in terms of its unavailability. This takes inspiration from the robust contingency-constrained unit commitment proposed in \cite{street2011contingency}.
    \item We exhaustively compare the proposed robust optimization problem with its deterministic counterpart.
\end{itemize}

The remainder of this paper is organized as follows. Section \ref{sec:formulation} presents the proposed day-ahead model for the aggregator of EVs in a residential area. Section \ref{sec:methodology} describes the methodology to transform the original problem into a single-level mixed-integer linear equivalent. Section \ref{sec:deterministic} provides the deterministic formulation for the aggregator's problem. Section \ref{sec:case} presents the case study and discusses the results. Finally, some relevant conclusions are duly drawn in Section \ref{sec:conclusion}.

\section{Problem Formulation}
\label{sec:formulation}
We consider an aggregator of EVs in a residential district. The usual pattern of this pool of EVs during weekdays is to depart from home to work in the early morning and to arrive home in the evening. The aggregator of EVs aims to minimize their total operation costs and it can be mathematically expressed as follows:
\begin{align}
&\min_{\Xi^{RB}} \hspace{3pt} \sum_{v \in \mathcal{V}} \sum_{t \in \mathcal{T}} \lambda_{t} c_{v,t} \Delta t + \sum_{v \in \mathcal{V}} \sum_{t \in \mathcal{T}}  P \left( s^+_{v,t} + s^-_{v,t}\right)  \label{robust_of}\\
&\text{subject to:}\notag\\
& e_{v,t} = e_{v,t-1} + \Delta t \eta_{v} c_{v,t} \alpha_{v,t} - \widehat{\xi}_{v,t} + s^+_{v,t} - s^-_{v,t} ,\notag\\
& \quad \forall t \in \mathcal{T}, v \in \mathcal{V} \label{robust_eq1}\\
& c_{v,t} \leq \overline{C}_{v} , \quad \forall v \in \mathcal{V}, t \in \mathcal{T} \label{robust_eq2}\\
& \underline{E}_{v} \leq e_{v,t}  \leq \overline{E}_{v}, \quad \forall v \in \mathcal{V}, t \in \mathcal{T} \label{robust_eq3}\\
& e_{v,N_T} = e_{v,0}, \quad \forall v \in \mathcal{V}  \label{robust_eq4}\\
& c_{v,t}, s^+_{v,t}, s^-_{v,t} \geq 0, \quad \forall v \in \mathcal{V}, t \in \mathcal{T} \label{robust_eq5}\\
& \xi^{wc}_{v} \geq  \sum_{t \in \mathcal{T}} \widehat{\xi}_{v,t} , \quad \forall v \in \mathcal{V}  \label{robust_eq6}\\
& \xi^{wc}_{v} = \min_{\alpha_{v,t}} \hspace{1pt} \left\{ \right. \sum_{t \in \mathcal{T}} \Delta t \hspace{1pt}  \eta_{v} c_{v,t} \alpha_{v,t}  \label{robust_eq7}\\
& \hspace{1.7cm}\sum_{t \in \mathcal{T}} \alpha_{v,t} \geq K_{v} : (\zeta_{v})  \label{robust_eq8}\\
& \hspace{1.7cm} \underline{\alpha}_{v,t} \leq \alpha_{v,t} \leq  \overline{\alpha}_{v,t} : (\underline{\beta}_{v,t}, \overline{\beta}_{v,t}),\quad \forall t \in \mathcal{T} \label{robust_eq9}\\
& \hspace{1.7cm}  \left.\right\}, \forall v \in \mathcal{V}, \notag
\end{align}

\noindent where the set of decision variables is $\Xi^{RB} = (c_{v,t}, e_{v,t}, s^+_{v,t}, s^-_{v,t}, \alpha_{v,t}, \xi^{wc}_{v})$.

The robust optimization model is driven by the minimization of the total costs \eqref{robust_of}, which comprise two terms: (i) the cost of the energy bought by the aggregator in the day-ahead market, and (ii) the penalty costs when the equation associated with the energy state-of-charge evolution is violated. This objective function is subject to technical constraints. Expressions \eqref{robust_eq1} model the energy state-of-charge evolution of the EV $v$ at time period $t$, while taking into account the energy required for transportation. These expressions include the variable $\alpha_{v,t}$ to account for the availability or unavailability of the EVs. Constraint \eqref{robust_eq2} imposes the maximum charging rate for electric vehicle $v$ and time period $t$. The maximum and minimum bounds for the energy state-of-charge are set in constraints \eqref{robust_eq3}. Expressions \eqref{robust_eq4} enforce boundary conditions on the energy state-of-charge of the EVs. Constraints \eqref{robust_eq5} define the non-negativity character of the upper-level decision variables.

Constraint \eqref{robust_eq6} enforces that the total demand due to transportation must be satisfied throughout the time horizon for the worst-case scenario of the EVs' availability for each EV. This worst-case scenario is obtained by minimizing \eqref{robust_eq7}, i.e., the {\color{black} energy available for transportation to EV $v$} throughout the optimization horizon, over the uncertainty set of its availability, which is defined by \eqref{robust_eq8}--\eqref{robust_eq9}. Dual variables are shown in parentheses after a colon. Constraint \eqref{robust_eq8} sets the minimum number of time intervals in which the electric vehicle $v$ must be available throughout the time horizon. Constraints \eqref{robust_eq9} serve us to easily enforce the availability or unavailability of the electric vehicle $v$ in time period $t$ by setting $\underline{\alpha}_{v,t} = \overline{\alpha}_{v,t} = 1$ or $\underline{\alpha}_{v,t} = \overline{\alpha}_{v,t} = 0$, respectively. {\color{black} Parameters $K_v$, $\overline{\alpha}_{v,t}$, and $\underline{\alpha}_{v,t}$ are estimated based on historical records. More generally, though, these parameters should be tuned to reflect the risk-aversion attitude of the aggregator.}

\section{Methodology}
\label{sec:methodology}
The lower-level decision variables $\alpha_{v,t}$ are binary and the lower-level problems become non-convex. Note, however, that the constraint matrix of each lower-level problem \eqref{robust_eq7}--\eqref{robust_eq9} is totally unimodular {\color{black} (see conditions (iii)-(iv) in Thm. 19.3 of \cite{schrijver1998theory})}. Besides, we assume that the parameters $K_v, \underline{\alpha}_{v,t}, \overline{\alpha}_{v,t}$ are integer. {\color{black} Hence, we can relax the binary character of $\alpha_{v,t}$ while ensuring that there exists an optimal solution for which these variables take integer values. Thus,} the relaxed lower-level problems are linear programs and thus, the original bilevel problem is transformed into a single-level equivalent by using results from duality theory of linear programming \cite{luenberger}, i.e., we apply the so-called duality-based approach. In a nutshell, we replace each lower-level optimization problem with its primal feasibility constraints, its dual feasibility constraints, and the equality corresponding to the strong duality theorem. Note that $\xi^{wc}_v$ is replaced either by the primal or dual lower-level objective function. Finally, we add back the binary character of the decision variables $\alpha_{v,t}$.

\subsection{Dual Lower-level Problem}
\label{method:LLO}
The dual problem of \eqref{robust_eq7}--\eqref{robust_eq9} for each electric vehicle $v$ is:
\begin{align}
& \max_{\zeta_{v},\underline{\beta}_{v,t},\overline{\beta}_{v,t}} \hspace{0.3cm} K_{v} \zeta_{v}  + \sum_{t \in \mathcal{T}} \left(\underline{\alpha}_{v,t} \underline{\beta}_{v,t}  + \overline{\alpha}_{v,t} \overline{\beta}_{v,t} \right)\label{robust_dual_of} \\
&\text{subject to:}\notag\\
& \zeta_{v} + \underline{\beta}_{v,t} + \overline{\beta}_{v,t} \leq \Delta t \hspace{1pt} \eta_{v} c_{v,t}, \quad \forall  t \in \mathcal{T} \label{robust_dual_eq1}\\
&  \underline{\beta}_{v,t} \geq 0, \overline{\beta}_{v,t} \leq 0, \quad \forall t \in \mathcal{T} \label{robust_dual_eq2}\\
&  \zeta_{v} \geq 0. \label{robust_dual_eq3}
\end{align}

\subsection{Robust Single-Level Formulation}
The resulting single-level robust formulation can be mathematically expressed as:

\begin{align}
&\min_{\Xi^{RS}} \hspace{3pt} \text{Objective function \eqref{robust_of}}  \label{robust_final_of}\\
&\text{subject to:}\notag\\
& \text{Constraints \eqref{robust_eq1}--\eqref{robust_eq5}} \label{robust_final_eq1}
\end{align}
\begin{align}
&  K_{v} \zeta_{v} + \sum_{t \in \mathcal{T}} \left(\underline{\alpha}_{v,t} \underline{\beta}_{v,t}  + \overline{\alpha}_{v,t} \overline{\beta}_{v,t} \right) \geq  \sum_{t \in \mathcal{T}} \widehat{\xi}_{v,t},  \forall v \in \mathcal{V} \label{robust_final_eq2} \\
& \text{Constraints \eqref{robust_eq8}--\eqref{robust_eq9}} \label{robust_final_eq3}\\
& \text{Constraints \eqref{robust_dual_eq1}--\eqref{robust_dual_eq3}} \label{robust_final_eq4}\\
& K_{v} \zeta_{v}  + \sum_{t \in \mathcal{T}} \left(\underline{\alpha}_{v,t} \underline{\beta}_{v,t}  + \overline{\alpha}_{v,t} \overline{\beta}_{v,t} \right) = \sum_{t \in \mathcal{T}} \Delta t \hspace{1pt} \eta_{v} c_{v,t} \alpha_{v,t},\notag\\
& \quad \forall v \in \mathcal{V} \label{robust_final_DS}\\
& \alpha_{v,t} \in \{0, 1\}. \label{robust_final_eq5}
\end{align}

Constraints \eqref{robust_final_eq1} are identical to the upper-level constraints \eqref{robust_eq1}--\eqref{robust_eq5}, constraints \eqref{robust_final_eq2} are identical to \eqref{robust_eq6}, lower-level primal and dual feasibility constraints are given in \eqref{robust_final_eq3}--\eqref{robust_final_eq4}, expression \eqref{robust_final_DS} is the equality associated with the strong duality theorem, and the binary character of $\alpha_{v,t}$ is imposed in \eqref{robust_final_eq5}. Problem \eqref{robust_final_of}--\eqref{robust_final_eq5} is characterized as a nonlinear program due to nonlinear products of continuous and binary variables (i.e., $c_{v,t} \alpha_{v,t}$) in constraints \eqref{robust_eq1} and \eqref{robust_final_DS}. By using integer algebra results \cite{floudas1995}, we can replace those constraints with the following set of linear ones:
%
\begin{align}
& e_{v,t} = e_{v,t-1} + \Delta t \eta_{v} z_{v,t} - \widehat{\xi}_{v,t} + s^+_{v,t} - s^-_{v,t} ,\notag\\
& \quad \forall t \in \mathcal{T}, v \in \mathcal{V} \label{robust_lin1}\\
& K_v \zeta_{v}  + \sum_{t \in \mathcal{T}} \left(\underline{\alpha}_{v,t} \underline{\beta}_{v,t}  + \overline{\alpha}_{v,t} \overline{\beta}_{v,t} \right) = \sum_{t \in \mathcal{T}} \Delta t \hspace{1pt} \eta_{v} z_{v,t},\notag\\
& \quad \forall v \in \mathcal{V} \label{robust_lin2}\\
& 0 \leq c_{v,t} - z_{v,t} \leq \left( 1 - \alpha_{v,t} \right) \overline{C}_v, \quad \forall t \in \mathcal{T}, v \in \mathcal{V} \label{robust_lin3}\\
& 0 \leq z_{v,t} \leq \alpha_{v,t} \overline{C}_v, \quad \forall t \in \mathcal{T}, v \in \mathcal{V}. \label{robust_lin4}
\end{align}

Note that the set of decision variables $\Xi^{RS} = (c_{v,t}, e_{v,t}, z_{v,t}, s^+_{v,t}, s^-_{v,t}, \alpha_{v,t}, \overline{\beta}_{v,t}, \underline{\beta}_{v,t}, \zeta_{v})$. The single-level mixed-integer linear robust optimization problem is hereinafter referred to as RO-EV.

\section{Deterministic Formulation}
\label{sec:deterministic}
The proposed robust optimization problem RO-EV is compared with its deterministic counterpart, namely DO-EV. The deterministic problem DO-EV can be expressed as follows:

\begin{align}
&\min_{\Xi^{DO}} \hspace{3pt} \hspace{3pt} \sum_{v \in \mathcal{V}} \sum_{t \in \mathcal{T}} \lambda_{t} c_{v,t} \Delta t + \sum_{v \in \mathcal{V}} \sum_{t \in \mathcal{T}}  P \left( s^+_{v,t} + s^-_{v,t}\right)  \label{do_of}\\
&\text{subject to:}\notag\\
%
& e_{v,t} = e_{v,t-1} + \Delta t \eta_{v} c_{v,t} - \widehat{\xi}_{v,t} + s^+_{v,t} - s^-_{v,t} ,\notag\\
& \quad \forall t \in \mathcal{T}, v \in \mathcal{V} \label{do_eq1}\\
& c_{v,t} \leq \overline{C}_{v} \widehat{\alpha}_{v,t} , \quad \forall v \in \mathcal{V}, t \in \mathcal{T} \label{do_eq2}\\
& \underline{E}_{v} \leq e_{v,t}  \leq \overline{E}_{v}, \quad \forall v \in \mathcal{V}, t \in \mathcal{T} \label{do_eq3}\\
& e_{v,N_T} = e_{v,0}, \quad \forall v \in \mathcal{V}  \label{do_eq4}\\
& c_{v,t}, s^+_{v,t}, s^-_{v,t} \geq 0, \quad \forall v \in \mathcal{V}, t \in \mathcal{T}, \label{do_eq5}
\end{align}

\noindent where the set of decision variables $\Xi^{DO} = (c_{v,t}, e_{v,t}, s^+_{v,t}, s^-_{v,t})$ and parameters $\widehat{\alpha}_{v,t}$ and $\widehat{\xi}_{v,t}$ are expected values. The computation of these parameters is described in Section \ref{sec:case}.

\section{Case Study}
\label{sec:case}
We assume a residential aggregator with 100 EVs. For the sake of simplicity, the technical parameters associated with each EV are identical: The maximum charging rate is 7.4 kW, the round-trip efficiency is 0.95, the minimum and maximum energy rates are 10 and 51 kWh, in that order, and the energy rating per kilometer is 0.137 kWh/km \cite{Technical_ZOE}.

\begin{figure}[t]
\centerline{\includegraphics[width=9.5cm]{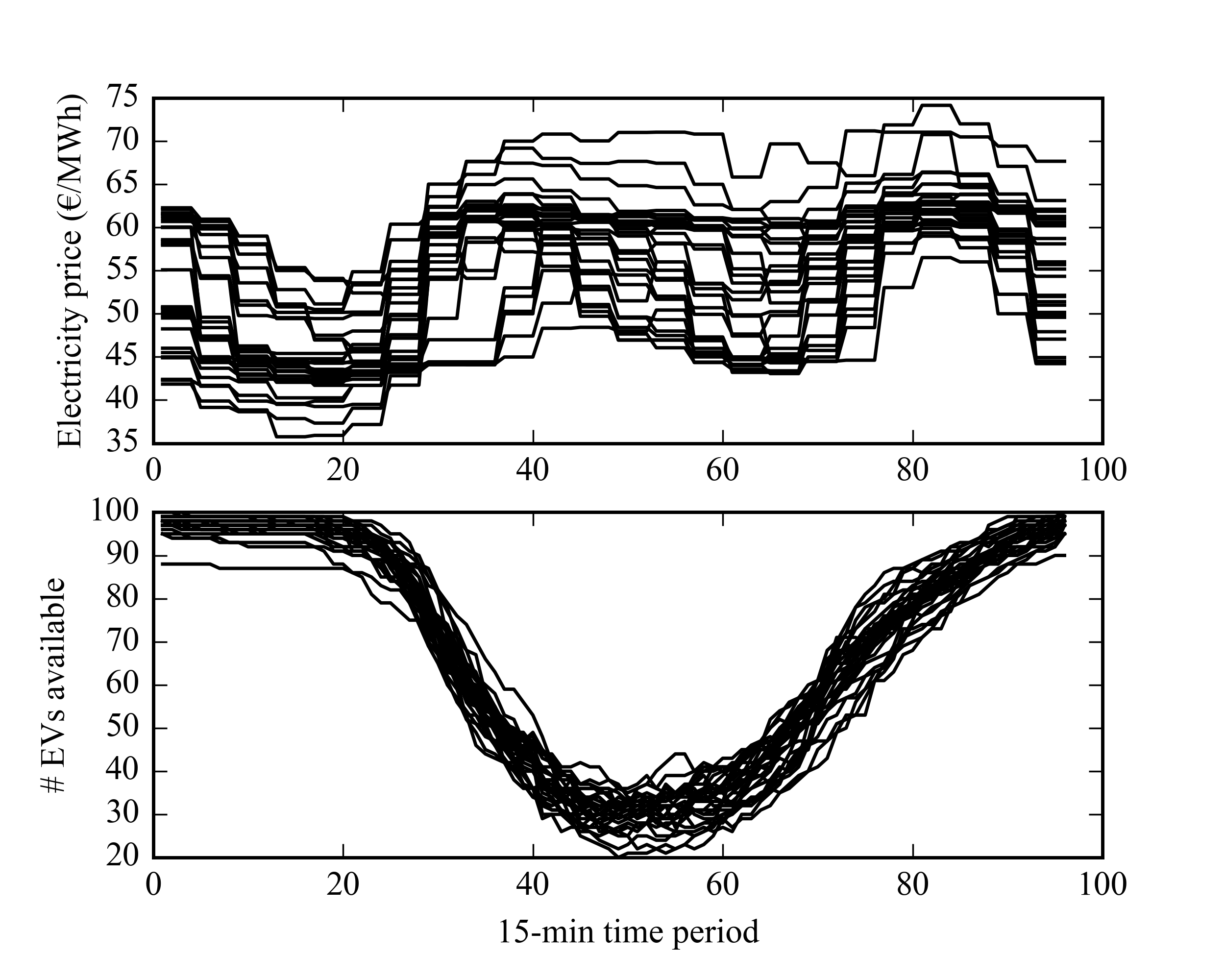}}
\vspace{-0.6cm}
\caption{Electricity prices (upper plot) and total number of EVs available (lower plot).}
\label{fig_data_prices}
\end{figure}

\begin{figure}[t]
\centerline{\includegraphics[width=8.5cm]{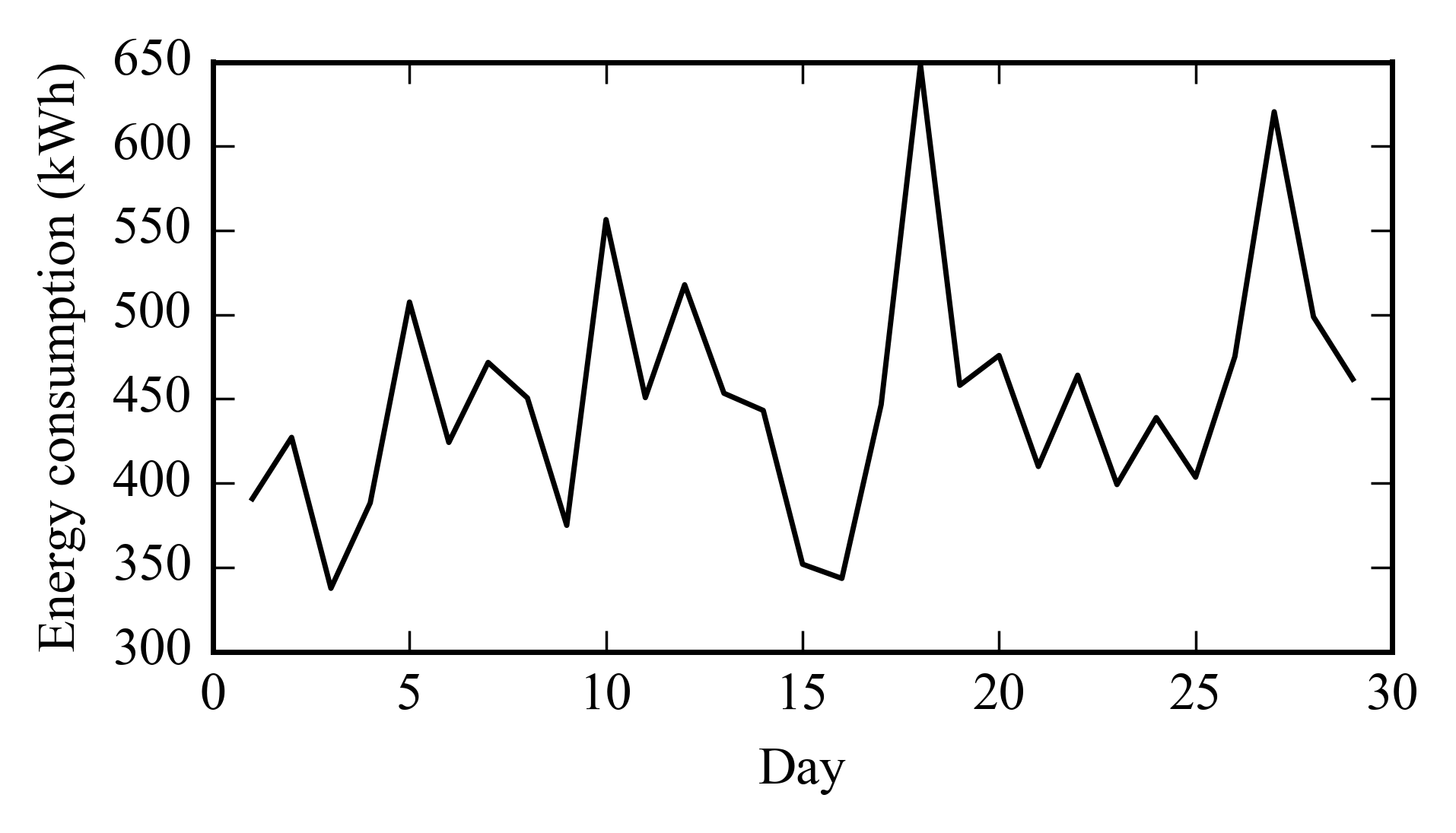}}
\vspace{-0.6cm}
\caption{Total energy consumption of the fleet of EVs.}
\label{fig_data_consumption}
\end{figure}

Due to the lack of real-life data about the parameters associated with the driving patterns (availability profiles and energy required for transportation) of EVs, we resort to the National Household Travel Survey (NHTS) 2017 \cite{NHTS}. This data base contains synthetic data about vehicle trips, miles and duration per trip, etc. From this data base, we can extract the availability status by using the departure/arrival time periods for each daily trip. Specifically, we assume that the EV is available until it begins its first daily trip and after it returns from its last daily trip for each day of the year. Otherwise the EV is unavailable and thus it may be in a motion status. The energy required for transportation $\xi_{v,t}$ can be computed as the product of the travelled distance and energy rating per kilometer (i.e., 0.137 kWh/km). The lower plot of Fig. \ref{fig_data_prices} shows the total number of EVs available in 15-min time periods for 29 days, whereas the Fig. \ref{fig_data_consumption} represents the energy consumption of the fleet of EVs per day. These data are used as realizations when validating the results in a real-time framework.

Fig. \ref{fig_data_prices} also represents the 15-min electricity prices, which can be obtained from the ENTSO-e Transparency Platform \cite{ENTSOE} in year 2018 in Spain. We also assume that the penalty term $P$ is set to 1000 $\textup{\euro}$/kWh. We run daily simulations with 15-min time steps for one month (specifically 29 days).

Note that based on the data provided in figures \ref{fig_data_prices} and \ref{fig_data_consumption}, we compute the expected availability $\widehat{\alpha}_{v,t}$, energy consumption $\widehat{\xi}_{v,t}$, and value of $K_v$ as follows: the expected values for day $D$, assuming that it is Monday, are the average over the data on the previous four Mondays. In addition, we assume that the day-ahead electricity prices $\lambda_t$ on day $D$ are the average over the last four days.

The simulations have been performed on a Windows 10 server with one CPU clocking at 2.8 GHz, 6 cores and 8 GB of RAM using CPLEX 12.6.3 \cite{cplex} under Pyomo 5.2 \cite{pyomo}. The optimality gap is set to 0\%.

\subsection{Day-ahead Operation of an Individual Electric Vehicle}
We first analyze the day-ahead operation for two individual EVs in day 21. We can identify two types of EVs depending on their driving patterns' behaviour: one whose daily routine is highly predictable and thus the departure/arrival times may be narrowed down to a few time periods (i.e. in this case the availability can be fixed in most time periods throughout the time horizon); and another one whose daily routine is unpredictable and thus its uncertainty set is larger.

\begin{figure}[tbp]
\centerline{\includegraphics[scale=0.38]{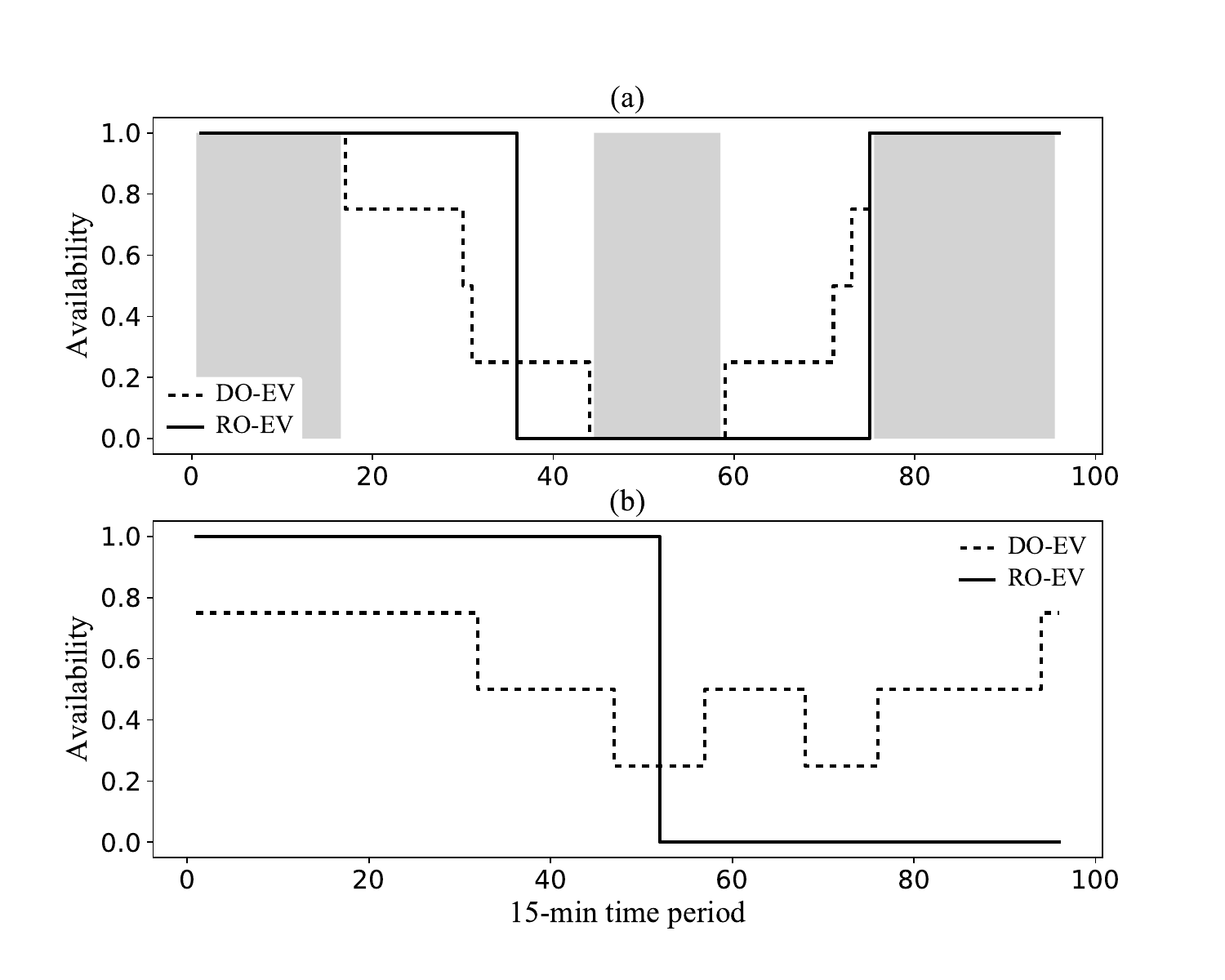}}
\vspace{-0.6cm}
\caption{Availability for the DO-EV and RO-EV for two EVs: (a) EV-A whose pattern is predictable to some extent, and (b) EV-B whose pattern is not predictable.}
\label{fig1}
\end{figure}

\begin{figure}[tbp]
\centerline{\includegraphics[scale=0.38]{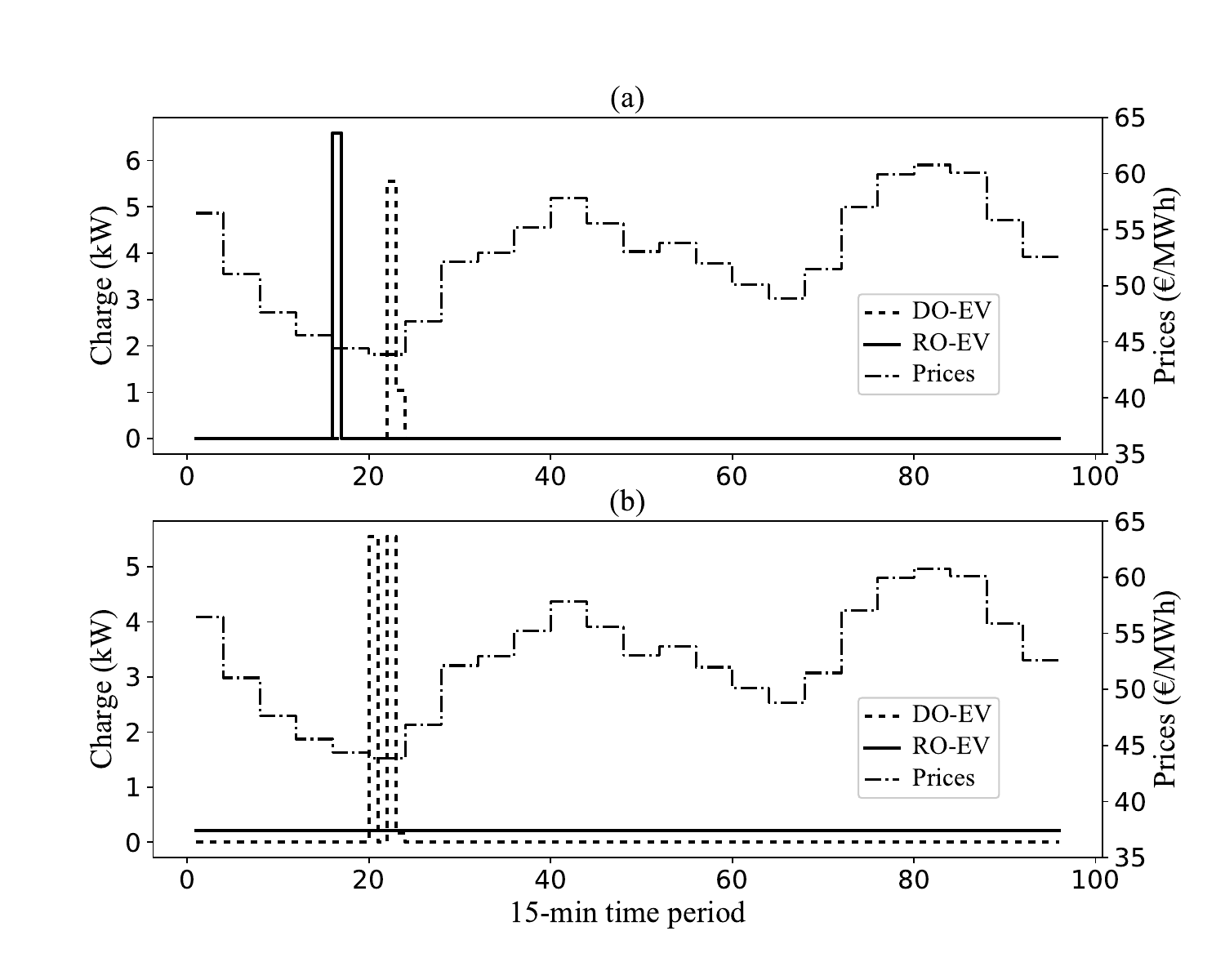}}
\vspace{-0.6cm}
\caption{Charging schedule for the DO-EV and RO-EV (left y-axis) and day-ahead electricity prices (right y-axis) for (a) EV-A and (b) EV-B.}
\label{fig2}
\end{figure}

\begin{figure}[tbp]
\centerline{\includegraphics[scale=0.38]{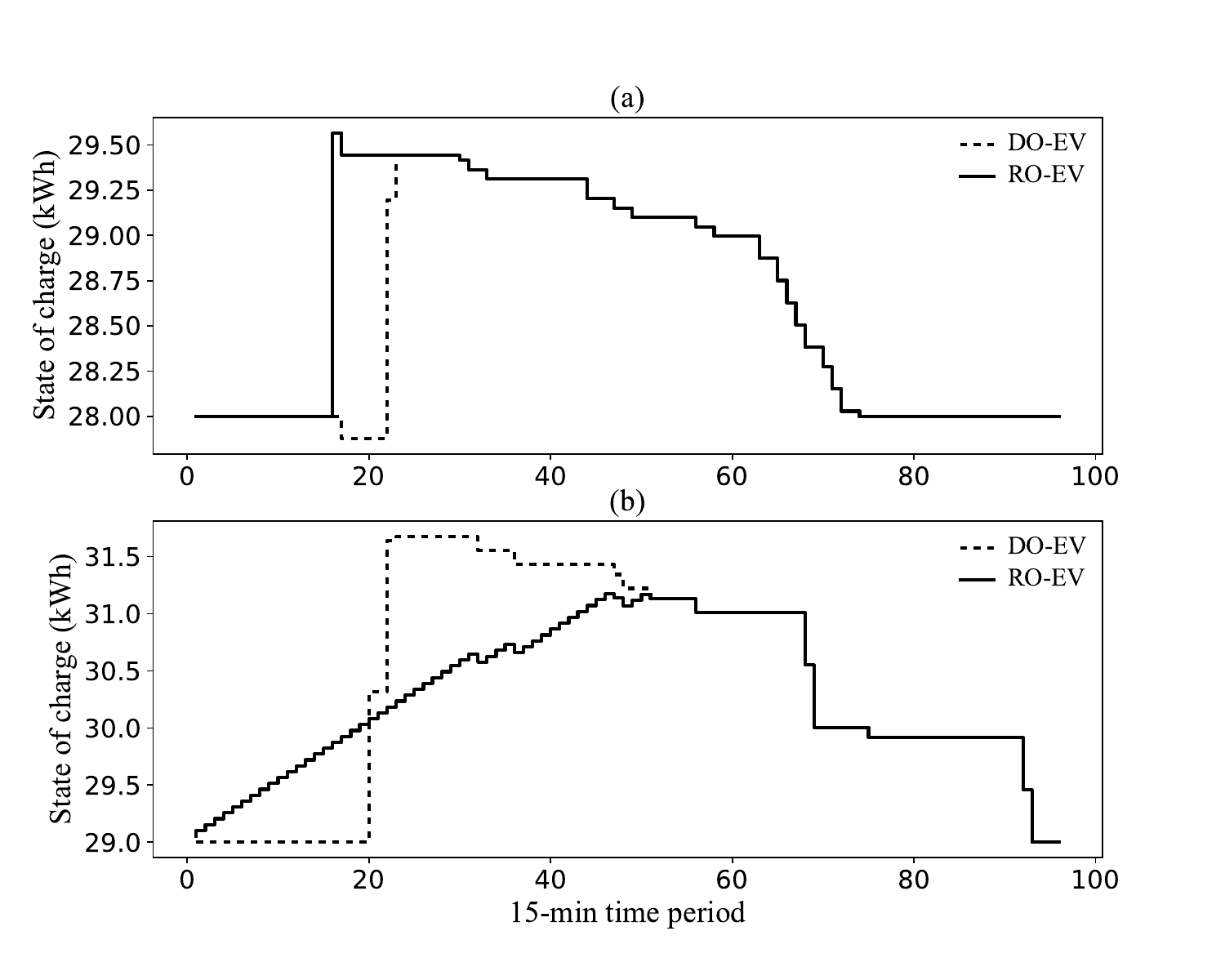}}
\vspace{-0.6cm}
\caption{Energy state-of-charge evolution for the DO-EV and RO-EV and for (a) EV-A and (b) EV-B.}
\label{fig3}
\end{figure}

Fig. \ref{fig1} shows the availability of two EVs, namely EV-A and EV-B, for both methods DO-EV and RO-EV. Note that, on the one hand, the expected availability for the DO-EV may be different from 0 or 1 because we average over historical data, and, on the other hand, the availability for the RO-EV is fixed in those time periods marked with a grey fill. The upper plot represents the availability of EV-A, whose pattern is predictable to some extent, i.e., the availability is known in the grey time intervals. The lower plot represents the EV-B, whose pattern is not predictable and thus, no grey zones are shown in the figure. Note also that parameter $K_v$ is set to 57 and 52, respectively, which means that at least 57 and 52 time periods the EV should be available. Figures \ref{fig2} and \ref{fig3} provide the charging schedule, day-ahead electricity prices, and energy state-of-charge, respectively.

In Fig. \ref{fig2}, we can observe that, with the RO-EV, the EV-A charges in time period 17, i.e., in the zone when the availability is certainly known and the price is significantly low, whereas the DO-EV schedules the charging of EV-A in time periods 23 and 24, i.e., a few time periods later when the prices are the lowest. In Fig. \ref{fig3}, we can see that the evolution of the energy state-of-charge is decreasing until the boundary condition is fulfilled. However, for the EV-B, we can observe a completely different behaviour since there is not any interval when the EV availability is known. Therefore, the RO-EV leads to a solution in which the EV charges equally throughout the whole optimization horizon, as can be observed in Fig. \ref{fig2}.(b). Therefore, the energy state of charge  is smoothly increasing during the first half of the day (see Fig. \ref{fig3}).


\subsection{Operation of the Fleet of Electric Vehicles}
In order to fairly compare the performance of both methods, RO-EV and DO-EV, we solve the real-time operation of the aggregator for one month by fixing the energy bought from the day-ahead market (i.e., $\sum_{v \in \mathcal{V}} c_{v,t}$) and assuming the realized values of the parameters associated with each EV. The goal of the real-time problem is to minimize the deviations from the energy balance in the EVs' batteries for the fleet of EVs, which are penalized with 1000 $\textup{\euro}$/kWh. As a result, we can compare the following metrics: day-ahead cost $C^{DA}$, day-ahead purchased power $P^{DA}$, and total deviations from the energy balance $D^{RT}$ in real-time, for each method. Both methods attain the optimal solution for one day in less than a minute.

Table \ref{tab2:results_day21} summarizes the results for day 21, which is the same day chosen in the previous section. The day-ahead operation cost of the aggregator increases by 5.8\% when using the RO-EV compared to the results from the DO-EV. In terms of total purchased energy, the RO-EV leads to an increase of 6.0\% with respect to the deterministic optimization. Finally, this excess of purchased energy causes that there are less deviations with the RO-EV than the DO-EV; specifically, the energy deviations are reduced by approximately 60\%.

\begin{table}[tbp]
\caption{Results for Day 21: Day-ahead Cost, Purchased Power, and Energy Deviations}
\vspace{-0.4cm}
\begin{center}
\begin{tabular}{|c|c|c|c|}
 \hline
 \multicolumn{2}{|c|}{\textbf{Metric}} &    \textbf{DO-EV}      &        \textbf{RO-EV}\\
\hline
\multicolumn{2}{|c|}{$C^{DA}$ (\euro)} & 31.0 & 32.8 \\
\hline
\multicolumn{2}{|c|}{$P^{DA}$ (kW)} & 1745.6 & 1851.2 \\
\hline
\multicolumn{2}{|c|}{$D^{RT}$ (kWh)} & 305.7 &123.4    \\
\hline
\end{tabular}
\label{tab2:results_day21}
\end{center}
\end{table}

\subsection{Monthly Operation of the Fleet of Electric Vehicles}
\begin{figure}[tbp]
\centerline{\includegraphics[width=9.5cm]{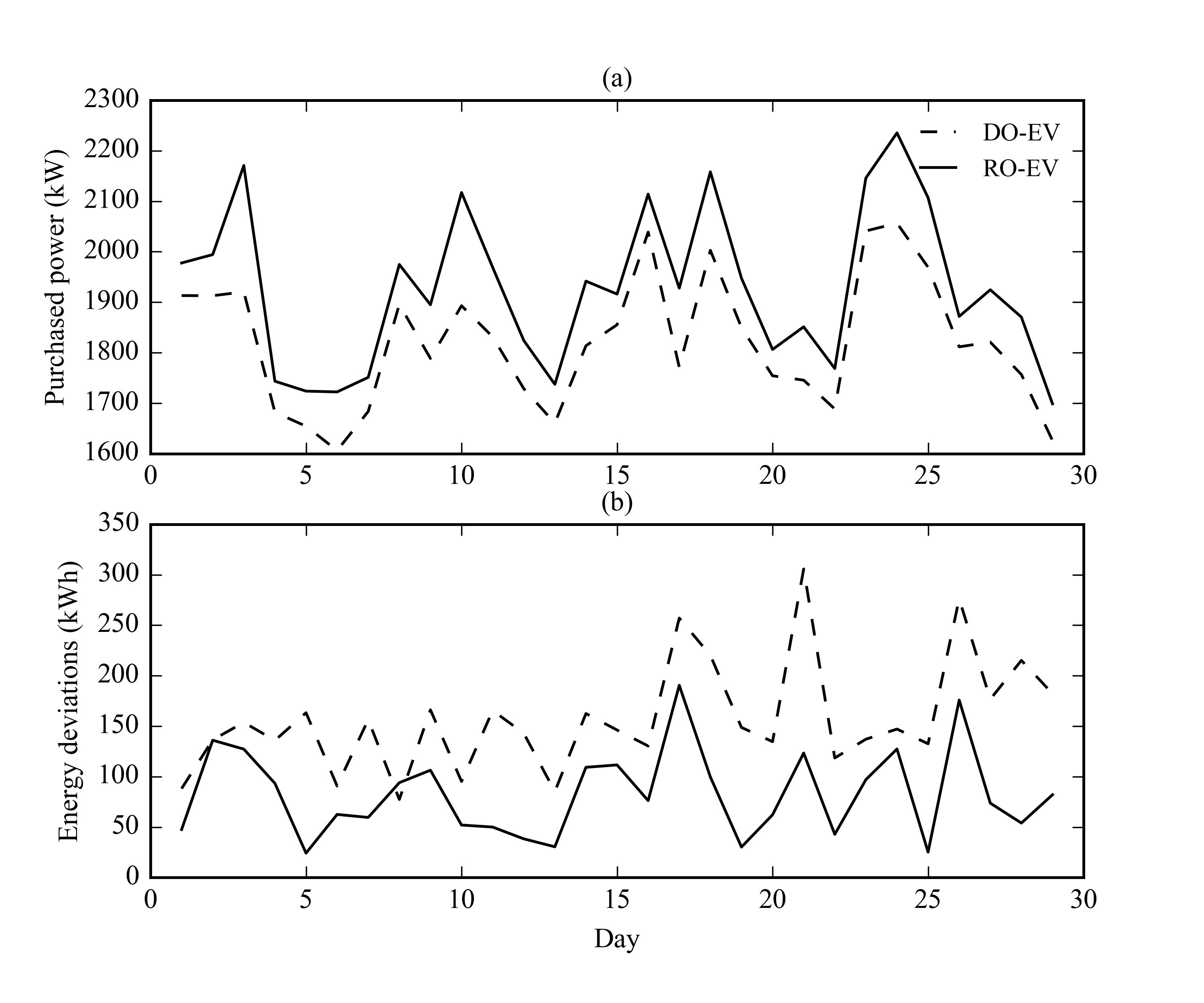}}
\vspace{-0.6cm}
\caption{Monthly results: (a) Day-ahead power purchased per day and (b) real-time energy deviations per day.}
\label{fig4}
\end{figure}

Next, we compare the performance of both methods for one month. Fig. \ref{fig4} represents the day-ahead purchased power and the real-time energy deviations for both methods and for each day of the month. Table \ref{tab1:results_month} provides the monthly results in terms of day-ahead cost and total energy deviations. The optimal results for a month are attained after 340.4 s with the DO-EV and 619.8 s with the RO-EV.

As can be seen in Table \ref{tab1:results_month}, although the day-ahead cost increases by 9.6\% when using the robust optimization compared to the results from the DO-EV, the total energy deviations decreased to almost half. We can also observe that the maximum, mean, and minimum values are also reduced when using the RO-EV. The cost increase is because the RO-EV schedules higher charging rates than the DO-EV to cope with the uncertainty on the availability and thus, the power purchased in the day-ahead market increases between 3.0\% for day 20 and 13.0\% for day 3 with respect to the power bought when using the DO-EV (see Fig. \ref{fig4}.(a)). As a result, the deviations of the energy balance of the vehicles' batteries are decreased all days except for day 8 and they may attain a reduction up to 83.0\%, as observed for day 5 in Fig.~\ref{fig4}.(b).

\begin{table}[tbp]
\caption{Monthly Results: Day-ahead Cost and Energy Deviations}
\vspace{-0.4cm}
\begin{center}
\begin{tabular}{|c|c|c|c|}
 \hline
 \multicolumn{2}{|c|}{\textbf{Metric}} &    \textbf{DO-EV}      &        \textbf{RO-EV}\\
\hline
\multicolumn{2}{|c|}{$C^{DA}$ (\euro)} & 586.6 & 643.0 \\
\hline
\multirow{4}{*}{$D^{RT}$ (kWh)} & Max. & 305.7 &190.5    \\
&Mean              &      156.9 &  82.9 \\
&Min.   &        77.5 &       24.3   \\
&Total &       4548.9 &      2404.3 \\
\hline
\end{tabular}
\label{tab1:results_month}
\end{center}
\end{table}

\section{Conclusion}
\label{sec:conclusion}
This paper provides a novel and computationally efficient model for the day-ahead operation of an aggregator of EVs in a residential district in which the energy bought is minimized subject to the technical characteristics of each EV. As a salient feature, the uncertainty on the availability of an EV is modelled via robust optimization so that the total energy required for transportation throughout the optimization horizon must be satisfied. In this problem, the uncertainty set leads to a mixed-integer linear program in the lower level and the duality-based approach can be applied thanks to the unimodularity of the system matrix and the convexity of the uncertainty set.

The case study reveals that the robust optimization model could lead to a substantial reduction of the daily deviations from the energy balance of the vehicles' batteries up to approximately 80\% compared to the results from a deterministic approach. Logically, this benefit comes at the expense of increasing the daily purchase costs in the day-ahead market around 15\% compared to the deterministic counterpart. {\color{black} Further research will be devoted to refining the uncertainty set of the driving patterns, conducting a sensitivity analysis to the parameters defining that set and exploring the scalability of the proposed model to fleets comprising thousands of EVs. Moreover, we plan to extend our approach to EVs capable of discharging power to the grid.}

\bibliographystyle{ieeetr}
\bibliography{mendeley}

\end{document}